\documentclass{wsci}

\begin{document}

\newcommand{\trh}{T_{\rm rh}}
\newcommand{\teff}{T_{\rm eff}}
\newcommand{\delcp}{\delta_{_{\rm CP}}}
\newcommand{\mh}{m_{_{\rm H}}}
\newcommand{\mw}{m_{_{\rm W}}}
\newcommand{\alphaw}{\alpha_{_{\rm W}}}
\newcommand{\ncs}{N_{_{\rm CS}}}

\title{Electroweak baryogenesis from preheating}

\author{Juan Garcia-Bellido}

\address{Theoretical Physics Group, Blackett Laboratory,
Imperial College, \\
Prince Consort Road, London SW7 2BZ, United Kingdom
\\E-mail: bellido@ic.ac.uk}

\maketitle

\abstracts{The origin of the matter-antimatter asymmetry remains one of
the most fundamental problems of cosmology. In this talk I present a
novel scenario for baryogenesis at the electroweak scale, without the
need for a first order phase transition. It is based on the out of
equilibrium resonant production of long wavelength Higgs and gauge
configurations, at the end of a period of inflation, which induces a
large rate of sphaleron transitions, before thermalization at a
temperature below critical.}

\section{Introduction}

Everything we see in the universe, from planets and stars, to galaxies
and clusters of galaxies, is made out of matter, so where did the
antimatter in the universe go? Is this the result of an accident, a
happy chance occurrence during the evolution of the universe, or is it
an inevitable consequence of some asymmetry in the laws of nature?
Theorists tend to believe that the observed excess of matter over
antimatter, $\eta = (n_{\rm B}-n_{\bar{\rm B}})/n_\gamma \sim 10^{-10}$,
comes from tiny differences in their fundamental interactions soon after
the end of inflation. It is known since Sakharov that there are three
necessary conditions for the baryon asymmetry of the universe to
develop.\cite{sakharov} First, we need interactions that do not conserve
baryon number B, otherwise no asymmetry could be produced in the first
place. Second, C and CP symmetry must be violated, in order to
differentiate between matter and antimatter, otherwise B non-conserving
interactions would produce baryons and antibaryons at the same rate,
thus maintaining zero net baryon number. Third, these processes should
occur out of thermal equilibrium, otherwise particles and antiparticles,
which have the same mass, would have equal occupation numbers and would
be produced at the same rate. The possibility that baryogenesis could
have occurred at the electroweak scale is very appealing. The Standard
Model is baryon symmetric at the classical level, but violates B at the
quantum level, through the chiral anomaly. Electroweak interactions
violate C and CP through the irreducible phase in the
Cabibbo-Kobayashi-Maskawa (CKM) matrix, but the magnitude of the
violation is probably insufficient to account for the observed baryon
asymmetry.\cite{KRS} This failure suggests that there must be other
sources of CP violation in nature.  Furthermore, the electroweak phase
transition is certainly not first order and is probably too weak to
prevent the later baryon wash-out. In order to account for the observed
baryon asymmetry, a stronger deviation from thermal equilibrium is
required. An alternative proposal is that of
leptogenesis,\cite{leptogen} which may have occurred at much higher
energies, and later converted into a baryon asymmetry through
non-perturbative sphaleron processes at the electroweak scale.

Recently, a new mechanism for electroweak baryogenesis was
proposed,\cite{BGKS} based on the non-perturbative and out of
equilibrium production of long-wavelength Higgs and gauge configurations
via parametric resonance at the end of inflation.\footnote{A similar
idea, based on topological defects,\cite{KT} was proposed at the same
time.} Such mechanism occurs very far from equilibrium and can be very
efficient in producing the required sphaleron transitions that gave rise
to the baryon asymmetry of the universe, in the presence of a new
CP-violating interaction, without assuming that the universe ever went
through the electroweak phase transition.

\section{The hybrid model}

The new scenario\cite{BGKS} considers a very economical extension of the
symmetry breaking sector of the Standard Model with the only inclusion
of a singlet scalar field $\sigma$ that acts as an
inflaton.\footnote{This field is not necessarily directly related to the
inflaton field responsible for the observed temperature anisotropies in
the microwave background.} Its vacuum energy density drives a short
period of expansion, diluting all particle species and leaving an
essentially cold universe, while its coupling to the Higgs field $\phi$
triggers (dynamically) the electroweak symmetry breaking. After
inflation, the coherent inflaton oscillations induce explosive Higgs
production, via parametric resonance.\cite{KLS,GBL}

As a toy model, we consider a hybrid model of inflation at the
electroweak scale. The resonant decay of the low-energy inflaton can
generate a high-density Higgs condensate characterized by a set of
narrow spectral bands in momentum space with large occupation
numbers. The system slowly evolves towards thermal equilibrium while
populating higher and higher momentum modes. The expansion of the
universe at the electroweak scale is negligible compared to the mass
scales involved, so the energy density is conserved, and the final
reheating temperature $\trh$ is determined by the energy stored
initially in the inflaton field. For typical model parameters\cite{BGKS}
the final thermal state has a temperature below the electroweak scale,
$\trh \sim 70\ {\rm GeV} < T_c \sim 100$~GeV. Since $\trh <T_c$, the
baryon-violating sphaleron processes, relatively frequent in the
non-thermal condensate, are Boltzmann suppressed as soon as the plasma
thermalizes via the interaction with fermions.

\begin{figure}[t]
\centering
\hspace*{-5.5mm}
\epsfig{file=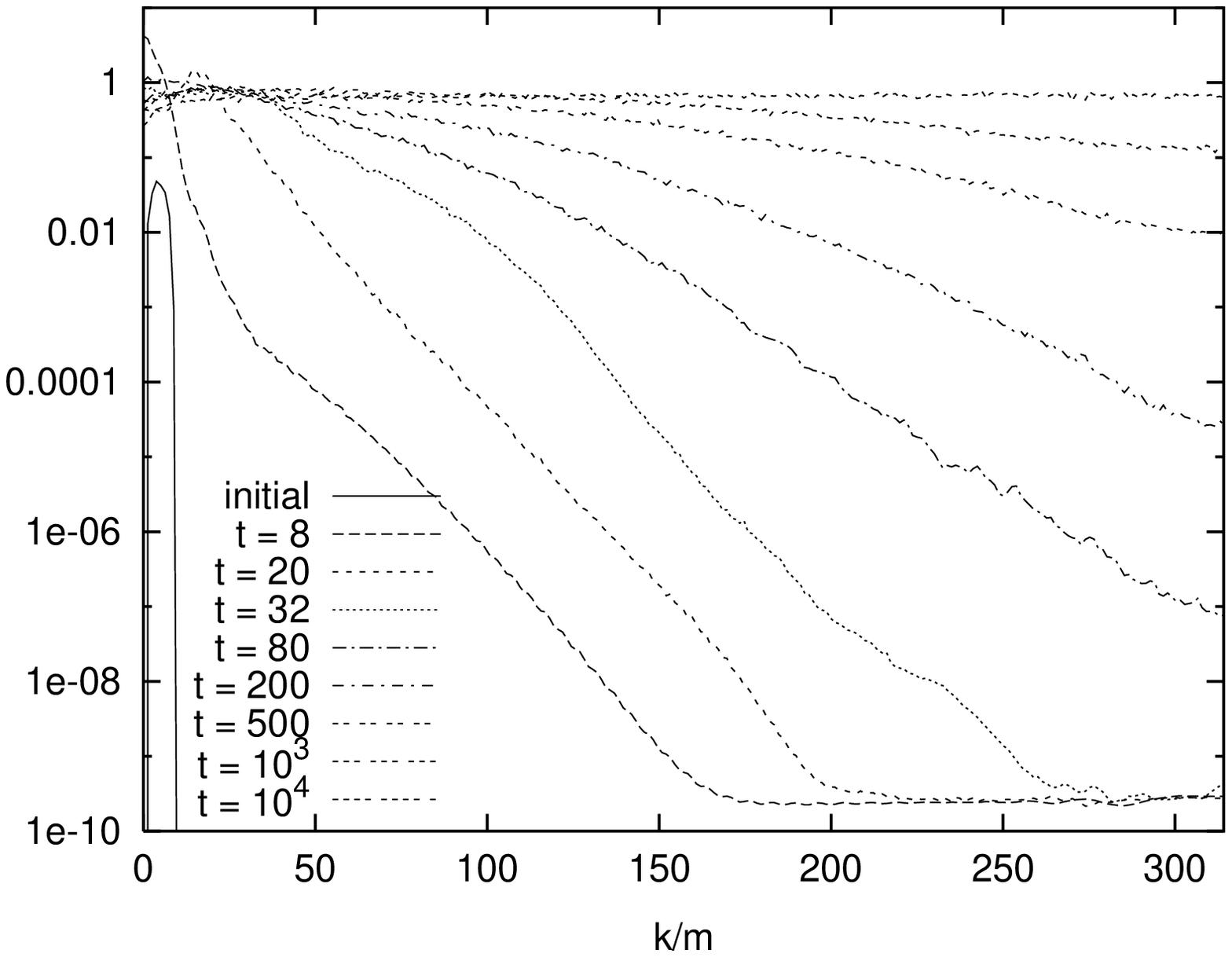, width=6cm}~~\epsfig{file=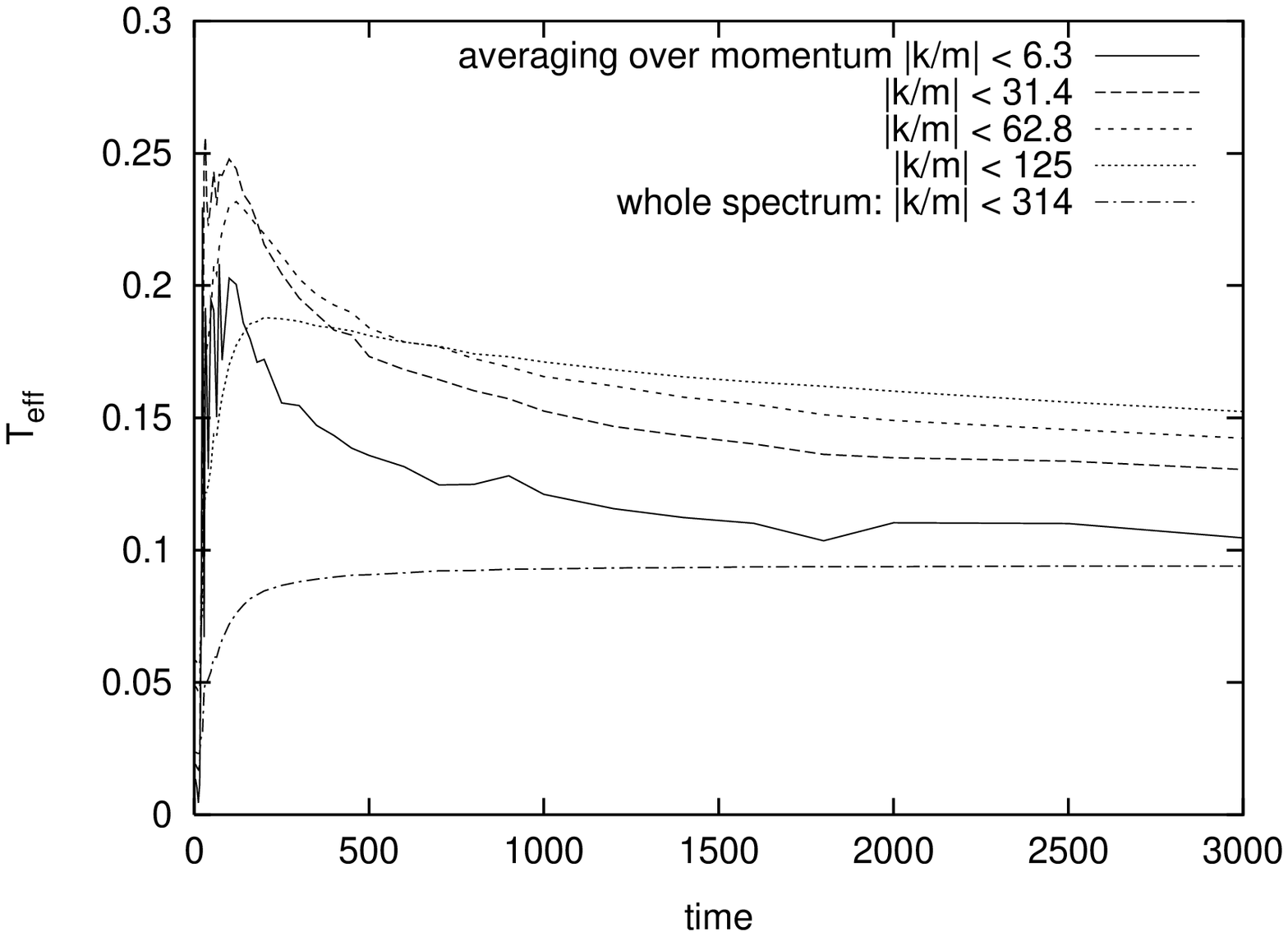, width=6cm}
\caption[fig1]{\label{fig1} The left panel shows the evolution of the
  Higgs spectrum $n_k\,\omega_k$, in units of $v=246$ GeV, from time 0
  to $10^4$ $v^{-1}$, as a function of momentum, $k/m$. The initial
  spectrum is determined by parametric resonance, and contains a set of
  narrow bands (solid line).  The subsequent evolution of the system
  leads to a redistribution of energy between different modes. Note how
  rapidly a ``thermal'' equipartition is reached for the long-wavelength
  modes. The right panel shows the time evolution of the effective
  temperature $\teff$ in units of $v$. Note the smooth rise and
  decline of the effective temperature with time.}
\end{figure}

One of the major problems that afflicted previous scenarios of
baryogenesis at the electroweak scale is the inevitability of a strong
wash-out of the generated baryons after the end of the CP-violation
stage during the phase transition. This problem was partially solved in
the new scenario,\cite{BGKS} where CP violation and efficient
topological (sphaleron) transitions coexist on roughly the same time
scale, during the resonant stage of preheating, while after-resonance
transitions are rapidly suppressed due to the decay of the Higgs and
gauge bosons into fermions and their subsequent thermalization below
100~GeV. For example, for the electroweak symmetry breaking VEV $v=246$
GeV, a Higgs self-coupling $\lambda \simeq 1$, and an inflaton-Higgs
coupling $g \simeq 0.1$, we find a negligible rate of expansion during
inflation, $H\simeq7\times 10^{-6}$ eV, and a reheating temperature
$\trh \simeq 70$ GeV. The relevant masses for us here are those in the
true vacuum, where the Higgs has a mass $\mh= \sqrt{2\lambda}\,v \simeq
350$ GeV, and the inflaton field a mass $m = gv \simeq 25$ GeV.  Such a
field, a singlet with respect to the Standard Model gauge group,
could be detected at future colliders because of its large coupling to
the Higgs field.

One of the most fascinating properties of rescattering after preheating
is that the long-wavelength part of the spectrum soon reaches some kind
of local equilibrium,\cite{TK} while the energy density is drained,
through rescattering and excitations, into the higher frequency modes.
Therefore, initially the low energy modes reach ``thermalization'' at a
higher effective temperature, see Fig.~\ref{fig1}, while the high energy
modes remain unpopulated, and the system is still far from true thermal
equilibrium. Thus, for the long wavelength modes, $n_k =
[\exp(\omega_k/T) -1]^{-1} \sim \teff/\omega_k \gg 1$, and the energy
per long wavelength mode is then $E_k \approx n_k\,\omega_k \approx
\teff$, or effectively equipartitioned. Since energy is conserved during
preheating, and only a few modes ($k\leq k_{\rm max} \sim 10\,m$) are
populated, we can compute the energy density in the Higgs and gauge
fields, to give,\cite{BGKS} in (3+1)-dimensions, $(10/6\pi^2)\teff
k_{\rm max}^3=\lambda v^4/4$, or
\begin{equation}\label{teff}
\teff \simeq 350\ {\rm GeV} \approx 5\,\trh\,.
\end{equation}
The temperature $\teff$ is significantly higher than the final reheating
temperature, $\trh$, because preheating is a very efficient mechanism
for populating just the long wavelength modes, into which a large
fraction of the original inflaton energy density is put. This means that
a few modes carry a large amount of energy as they come into partial
equilibrium among themselves, and thus the effective ``temperature'' is
high. However, when the system reaches complete thermal equilibrium,
the same energy must be distributed between all the modes, and thus
corresponds to a much lower temperature.

\section{Baryon asymmetry of the universe}

The Higgs and gauge resonant production induces out of equilibrium
sphaleron transitions. Sphalerons are large extended objects sensitive
mainly to the infrared part of the spectrum. We conjectured\cite{BGKS}
that the rate of sphaleron transitions at the non-equilibrium stage of
preheating after inflation could be estimated as $\Gamma_{\rm sph}
\approx \alphaw^4\teff^4$, where $\teff$ is the effective temperature
associated with the local ``thermalization'' of the long wavelength
modes of the Higgs and gauge fields populated during preheating.

In the Standard Model, baryon and lepton numbers are not conserved
because of the non-perturbative processes that involve the chiral
anomaly:
\begin{equation}
\partial_\mu j_{_B}^\mu = \partial_\mu j_{_L}^\mu =
{3g_{_W}^2\over32\pi^2}\,F_{\mu\nu}\tilde F^{\mu\nu}\,.
\end{equation}
Moreover, since sphaleron configurations connect vacua with different
Chern-Simons numbers, $\ncs$, they induce the corresponding changes
in the baryon and lepton number, $\Delta B = \Delta L = 3 \Delta \ncs$.

\begin{figure}[t]
\centering
\hspace*{-4.5mm}
\epsfig{file=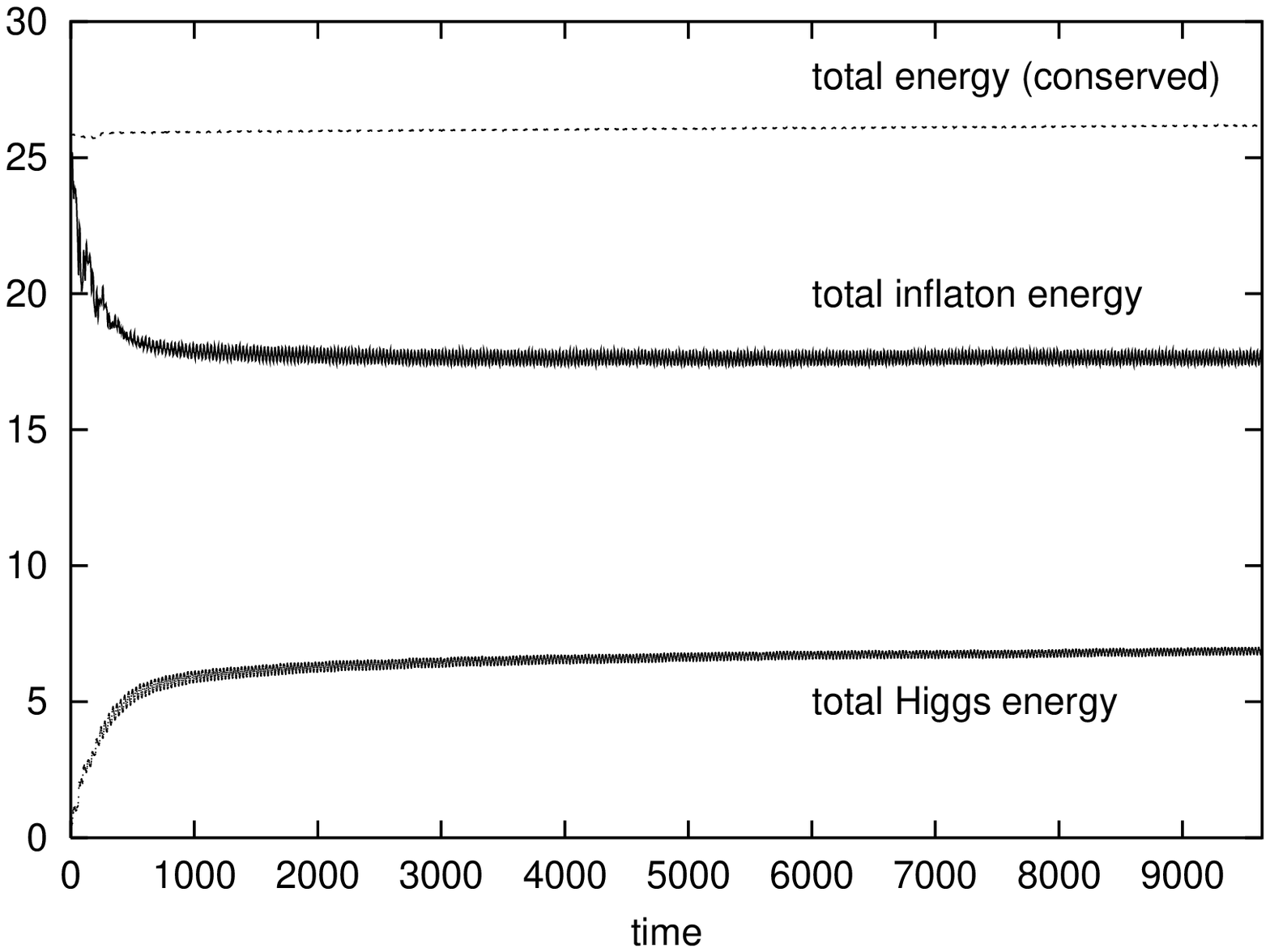, width=6cm}~~\epsfig{file=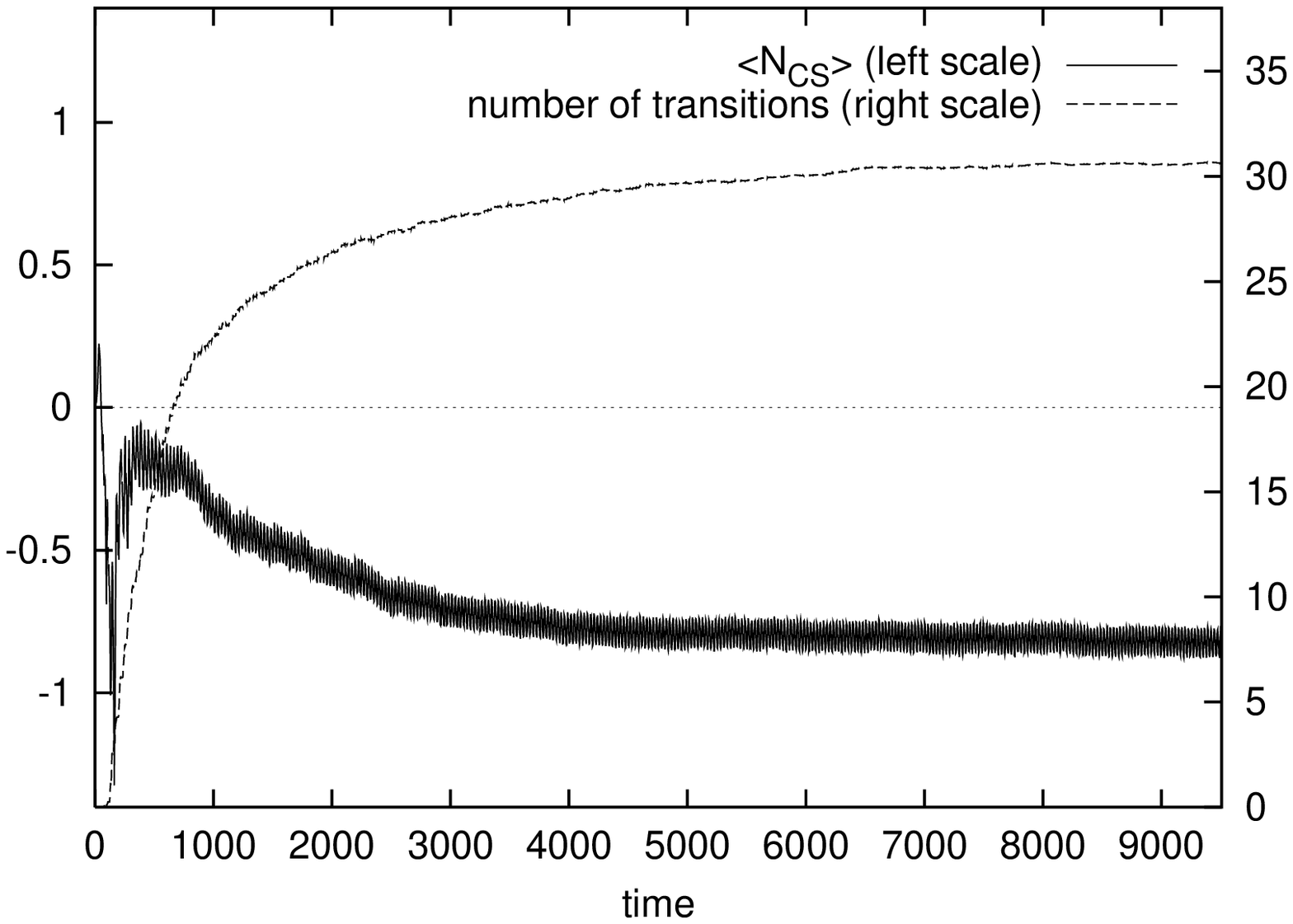, width=6cm}
\caption[fig2]{\label{fig2} The left panel shows the time evolution of
  the inflaton and Higgs energies in the case of a low energy
  resonance.\cite{GBG} The Higgs acquires here only about a third of
  the initial energy, while the inflaton zero-momentum mode retains the
  remaining two thirds. The right panel shows the continuous production
  of baryons as a result of correlations between the topological
  transition rate and the CP-violating operator in the Lagrangian. For a
  detailed description see Refs.~[4,9]. The solid line represents the
  shift in the Chern-Simons number, $\ncs$, averaged over an ensemble of
  a few hundred independent runs. The dashed line is the integral $\int
  \Gamma_{\rm sph}\,dt$, i.e. the average number of topological
  transitions accumulated per individual run.  Note the remarkable
  similarity of both curves for $t > 1000$. This means that all
  transitions at this stage are equally efficient in generating baryons,
  changing the Chern-Simons number by about $-1/20$ per transition for
  many oscillations, demonstrating the absence of baryon wash-out in the
  model.}
\end{figure}

A baryon asymmetry can therefore be generated by sphaleron transitions
in the presence of C and CP violation. There are several possible
sources of CP violation at the electroweak scale.  The only one
confirmed experimentally is due to CKM mixing of quarks that introduces
an irreducible CP-violating phase, which is probably too small to cause
a sufficient baryon asymmetry. Various extensions of the Standard Model
contain additional scalars (e.g. extra Higgs doublets, squarks,
sleptons, {\it etc.}) with irremovable complex phases that also lead to
C and CP violation.

We are going to model the effects of CP violation with an effective field
theory approach. Namely, we assume that, after all degrees of freedom
except the gauge fields, the Higgs, and the inflaton are integrated out,
the effective Lagrangian contains some non-renormalizable operators that
break CP. The lowest, dimension-six operator of this sort in (3+1)
dimensions is\cite{MS}
\begin{equation}\label{cpnonc}
{\cal O} = {\delcp\over M_{\rm new}^2}\phi^\dagger\phi \,
{3g_{_W}^2\over32\pi^2}\,F_{\mu\nu}\tilde F^{\mu\nu} \,.
\end{equation}
The dimensionless parameter $\delcp$ is an effective measure of CP
violation, and $M_{\rm new}$ characterizes the scale at which the new
physics, responsible for this effective operator, is important. Of
course, other types of CP violating operators are possible although,
qualitatively, they lead to the same picture.\cite{BGKS}

If the scalar field is time-dependent, the vacua with different
Chern-Simons numbers are not degenerate. This can be described
quantitatively in terms of an effective chemical potential, $\mu_{\rm
eff}$, which introduces a bias between baryons and
antibaryons,\cite{BGKS} $\mu_{\rm eff}\simeq \delcp\,{d\over dt}
\langle\phi^\dagger\phi\rangle/ M_{\rm new}^2$. Although the system
is very far from thermal equilibrium, we will assume that the evolution
of the baryon number $n_{\rm B}$ can be described by a Boltzmann-like
equation, where only the long-wavelength modes contribute, $\dot n_{\rm
B} = \Gamma_{\rm sph}\,\mu_{\rm eff}/\teff - \Gamma_{\rm B}\,n_{\rm B}$,
with $\Gamma_{\rm B} = (39/2) \Gamma_{\rm sph}/\teff^3 \sim
20\,\alphaw^4\teff$. The temperature $\teff$ decreases with time because
of rescattering, see Fig.~\ref{fig1}. The energy stored in the
low-frequency modes is transferred to the high-momentum modes.

The rate $\Gamma_{\rm B}$, even at high effective temperatures, is much
smaller than other typical scales in the problem. Indeed, for $\teff
\sim 400$ GeV, $\Gamma_{\rm B} \sim 0.01$ GeV, which is small compared
to the rate of the resonant growth of the Higgs condensate. It is also
much smaller than the decay rate of the Higgs into W's and the rate of W
decays into light fermions.  Therefore, the last term in the Boltzmann
equation never dominates during preheating and the final baryon
asymmetry can be obtained by integrating the Boltzmann equation
\begin{equation}
n_{\rm B} = \int dt\,\Gamma_{\rm sph}(t) {\mu_{\rm eff}(t)\over\teff(t)}
\ \simeq \ \Gamma_{\rm sph}\,{\delcp\over\teff}\,
{\langle\phi^\dagger\phi\rangle\over M_{\rm new}^2}\,,
\end{equation}
where all quantities are taken at the time of thermalization.
This corresponds to a baryon asymmetry
\begin{equation}\label{nB}
{n_{\rm B}\over s} \simeq {45\alphaw^4\delcp\over
2\pi^2\,g_*} {\langle\phi^\dagger\phi\rangle
\over M_{\rm new}^2}\,\Big({\teff\over\trh}\Big)^3\,,
\end{equation}
where $g_*\sim 10^2$ is the number of effective degrees of freedom that
contribute to the entropy density $s$ at the electroweak scale. Taking
$\langle\phi^\dagger\phi\rangle \simeq v^2=(246\,{\rm GeV})^2$, the
scale of new physics $M_{\rm new} \sim 1$ TeV, the coupling
$\alphaw\simeq 1/29$, the temperatures $\teff \simeq 350$ GeV and $\trh
\simeq 70$ GeV, we find
\begin{equation}\label{BAU}
{n_{\rm B}\over s} \simeq 3\times10^{-8}\,\delcp\,{v^2\over M_{\rm new}^2}
\Big({\teff\over\trh}\Big)^3 \ \simeq \ 2\times10^{-7}\,\delcp\,,
\end{equation}
consistent with observations for $\delcp \simeq 10^{-3}$, which is a
typical value from the point of view of particle physics beyond the
Standard Model. Therefore, baryogenesis at preheating can be very
efficient in the presence of an effective CP-violating operator coming
from some yet unkown physics at the TeV scale.

An important peculiarity of the new scenario is that it is possible for
the inflaton condensate to remain essentially spatially homogeneous for
many oscillation periods, even after the Higgs field has been produced
over a wide spectrum of modes. These inflaton oscillations induce a
coherent oscillation of the Higgs VEV through its coupling to the
inflaton, and thus induce CP-violating interactions arising from
operators (\ref{cpnonc}) containing the Higgs field. These oscillations
affect the sphaleron transition rate $\Gamma_{\rm sph}$ as well, since
the Higgs VEV determines the height of the sphaleron barrier, therefore
producing strong time correlations between variations in the rate
$\Gamma_{\rm sph}$ and the sign of CP violation.\cite{GBG} It is this
correlation between CP violation and the growth in the rate of sphaleron
transitions which ensures that the baryonic asymmetry generated is
completely safe from wash-out, because of the long-term nature of CP
oscillations. Depending on initial conditions, the rate $\Gamma_{\rm
sph}$ can finally vanish, e.g. due to the (bosonic) thermalization of
the Higgs field, as seen in Fig.~\ref{fig2}, but this doesn't affect the
continuous pattern of CP-$\Gamma_{\rm sph}$ correlations. In other
words, these correlations effectively give rise to a permanent and
constant CP violation, thus preventing the generated asymmetry from
being washed out.\cite{GBG}

\section*{Acknowledgments}
This research was supported by the Royal Society of London, through a
University Research Fellowship at Imperial College, and a Collaborative
Grant with Dimitri Grigoriev.

\end{document}